\documentclass[draft,reqno,12pt]{amsart}

\usepackage{amssymb}
\usepackage{enumerate}
\usepackage{a4wide}
 
\newcommand{\cl}{\mathcal{L}}

\newcommand{\E}{\mathbb{E}}

\newcommand{\R}{\mathbb{R}}

\newcommand{\V}{\mathbb{V}}

\newcommand{\X}{\mathbb{X}}

\newcommand{\Z}{\mathbb{Z}}

\let\a=\alpha

\let\d=\delta
\let\e=\epsilon
\let\ve=\varepsilon

\let\g=\gamma
\let\ve=\varepsilon
\let\l=\lambda

\let\n=\eta
\let\r=\rho
\let\t=\theta

\let\G=\Gamma
\renewcommand{\L}{\Lambda}

\newcommand{\1}{\,\rlap{\small 1}\kern.13em 1}

\newcommand{\sqr}[2]{{\vcenter{\hrule height.#2pt%
                      \hbox{\vrule width.#2pt height#1pt\kern#1pt%
                            \vrule width.#2pt}%
                      \hrule height.#2pt}}}
\newcommand{\cqfd}{\hfill$\mathchoice\sqr46\sqr46\sqr{1.5}2\sqr{1}2$\par}

\renewcommand{\limsup}{\mathop{\overline{\hbox{\rm lim}}}}

\newcommand{\vp}{\varphi}
\newcommand{\wh}{\widehat}
\newcommand{\wt}{\widetilde}

\let\bl=\bigl
\let\br=\bigr
\let\e=\ve

\newtheorem{thm}{Theorem}[section]
\newtheorem{lem}{Lemma}[section]
\newtheorem{prop}{Proposition}[section]

\begin{document}

\title[Hydrodynamics of a driven lattice gas]{Hydrodynamics of a
  driven lattice gas with open boundaries:  the asymmetric simple
  exclusion.} 

\author{O. Benois
   \and R. Esposito \and R. Marra \and M. Mourragui}
\address{{\rm O. Benois and M. Mourragui}\newline
Laboratoire de math\'ematiques Rapha\"el
     Salem, UMR 6085, Universit\'e de Rouen, 76821 Mont Saint Aignan,
     France
}
\address{{\rm R. Esposito}\newline
Centro interdisciplinare Linceo ``Beniamino Segre', via della Lungara 10, 00165 Roma, Italy; {\rm on
leave from} Dipartimento di Matematica Pura ed Applicata, Universit\`a di L'Aquila, 67100  Coppito, AQ,
Italy}
\address{{\rm R. Marra}\newline
Dipartimento di Fisica e Unit\`a INFM, 
Universit\`a di Roma Tor Vergata, Via della Ricerca Scientifica,
00133 Roma, Italy}

\begin{abstract}
  We consider the asymmetric simple exclusion process in $d\ge 3$ with
  open boundaries. The particle reservoirs of constant densities are
  modeled by birth and death processes at the boundary. We prove that,
  if the initial density and the densities of the boundary reservoirs
  differ for order of $\ve$ from $1/2$, the density empirical field,
  rescaled as $\e^{-1}$, converges to the solution of the
  initial-boundary value problem for the viscous Burgers equation in a
  finite domain with given density on the boundary.
  
\end{abstract}
\maketitle

\section{Introduction}
A driven lattice gas with open boundaries is a system of particles
jumping at random on a lattice, subject to the action of an external
field and exchanging matter with a reservoir at his boundary. The
combined action of the force field and the density gradient induced by
the boundary conditions forces the system to reach a stationary
non-equilibrium state.  Systems of this kind show a complex behavior
exhibiting non-equilibrium phase transitions \cite{K}, \cite{SZ}. The
simplest example of driven lattice gas is the asymmetric simple
exclusion (ASEP). In the one-dimensional totally asymmetric case an
explicit stationary solution is known \cite{D} showing a very reach
phase diagram with different behavior of the steady current depending
on the values of the fixed densities on the boundaries.

In this paper we study the time-dependent measure of the ASEP with
open boundaries in $d\ge 3$ in the macroscopic limit.  The system is
contained in a finite cylinder $\Lambda_\e=[-\ve^{-1},\ve^{-1}]\times
\pi_\ve^{d-1}$, with $\pi_\ve^{d-1}$ the $(d-1)$-dimensional
microscopic torus of size $2\ve^{-1}+1$ with the axes in the direction
$x_1$, namely we impose periodic boundary conditions in all the
directions but $x_1$. In the bulk particles jump to one of the nearest
neighbors if empty with jump rate $p_{e_i}$ ($p_{-e_i}$) in the
direction $e_i$ ($-e_i$). We assume that the vector of components
$\delta_i=p_{e_i}-p_{-e_i}$ is such that $\delta_1>0$. On the
boundaries $x_1=-\ve^{-1},\ve^{-1}$ we allow for production and
destruction of particles in the following way.  Let $b(u)$ be smooth
functions on $[-1,1]\times\pi_1^{d-1}$.  A particle is added
independently in each site of $x_1=-\ve^{-1}$, when the site is empty,
with rate $\d_1(1/2 +\ve b(\ve x))$ and removed independently in each
site of $x_1=\ve^{-1}$, when the site is occupied, with rate $\d_1(1/2
-\ve b(\ve x))$.  If $b(u)$ has constant values $b_{\pm}$, this choice
of the jump rates corresponds to coupling the system to reservoirs of
constant densities $1/2 +\e b_-$ and $1/2+\e b_+$ respectively. In the
paper we write the computations for this cylinder geometry, but the
proof could be extended to a more general convex domain. The initial
measure is the local equilibrium corresponding to a density profile
which is a perturbation of order $\ve$ of a constant profile:
$\rho_0(\ve x)=1/2 +\ve m_0 (\ve x)$.

We prove a law of large
numbers stating the weak convergence of the rescaled empirical field ($\eta_t(x)=0,1$ is the occupation number in the
site $x$ at time $t$)
\[
\ve^{d-1}\sum_{x\in\L_{\ve}} \Bigl(\n_{\ve^{-2 }t}(x)
-\frac{1}{2}\Bigr) \delta(\ve x)
\]
with $\d(\cdot)$ the Dirac measure, to the solution of the viscous
Burgers equation in the domain $\Lambda_1$ with density $b(u)$ on the
boundary $\Gamma$
\[
\left\{ \begin{array}{lll} \partial_t m (t,u) & = \delta\cdot\nabla_u
    \big( m(t,u)\big)^2
    +\sum_{ i,j=1}^d D_{i,j}\partial^2_{u_i,u_j}m(t,u) \\
    m(0,\,\cdot \,) & = m_0 (\,\cdot \,)\\
    m(t,\,\cdot \,){\Big|_\G} & =b(\,\cdot \,) \quad\text{for}\ t\ge 0
    ,\end{array} \right.\qquad\qquad
\]
where $D$ is a positive definite diffusion matrix whose expression is
given by the Green-Kubo formula for this model. We remind here that
the transport matrix is supposed to be infinite in $d\le 2$ \cite{Sp}.

This special choice of the initial condition as well as of the
boundary conditions (the difference between the top and the bottom
densities is of order $\e$) is forced by the fact that we want to
study the behavior of the system on the diffusive time scale $\e^{-2}$
to see the effect of a finite dissipation. On this time scale the
transport term, which involves first order time derivatives, is
enhanced by a factor of order $\e^{-1}$ so that if it is of order $\e$
at time $0$ its contribution stays finite in the limit. This is also
called incompressible limit \cite{EMY1} in analogy with the
Navier-Stokes case \cite{EMY2}.

There are few rigorous results on the hydrodynamic limit for
interacting particle systems in a bounded domain.  In \cite{ELS} it is
proved the hydrodynamic limit for one-dimensional gradient systems
both in the time-dependent and stationary case. In \cite{KLO} the
analogous result has been obtained in the stationary case for
one-dimensional non-gradient models. Finally, in \cite{LMS} the latter
result has been extended to the $d$-dimensional case. All these papers
deal with bulk reversible dynamics. If the lattice gas is driven by
the boundary conditions, in general the stationary measure does not
coincide with the invariant measure of the dynamics in the infinite
volume, so that even if the generator of the bulk dynamics is
reversible versus its invariant measure, the total dynamics is not.
The asymmetric simple exclusion is an example of a lattice gas driven
by an external field and it is not reversible w.r.t. his invariant
measure in the infinite volume, which is a product measure. The case
we consider in this paper is an example of a lattice gas driven by
both external field and boundary conditions, so that there are two
sources of non-reversibility. The effect of the asymmetry is seen in
the macroscopic current as the term $\delta_i m^2$ which gives rise to
the transport term in the macroscopic equation.  We have to face the
difficulty of non reversibility in the bulk and the fact that the
stationary measure is not explicitly known.

The first problem has been solved in [EMY1] where it is proved the law
of large numbers in the incompressible limit for the system without
reservoirs. The method used there is based on the use of relative
entropy of the true measure with respect to some suitable local
equilibrium measure and the main point is to prove that this relative
entropy vanishes in the limit $\e\to 0$. In this case it is natural to
assume for the local equilibrium measure a product measure because
this is the invariant measure of the generator. However, for the open
system it is not clear which is the good candidate for describing the
system for $\e$ small. We introduce a measure $\mu_\ve$ which is a
product measure with chemical potential $\e\lambda(t,\ve x)$ such that
$E^{\mu_\ve}[\eta_x]=1/2+\ve m(t,x)$, with $m$ solution of the
macroscopic equation. We prove that the relative entropy
$s(\mu|\mu_\ve)$ of the non-equilibrium measure $\mu$ w.r.t. $\mu_\ve$
satisfy
\begin{equation}
\lim_{\e\to 0}\e^{-2}s(\mu|\mu_\ve)=0 \label{eq:0}
\end{equation}
which implies the law of large numbers. One can understand the factor
$\e^{-2}$ noting that the specific entropy of $\mu_\ve$ is of order
$\e^2$.  To get this result, we introduce an auxiliary measure
$\tilde\mu_\e$ which differs from $\mu_\ve$ near the boundary and
prove the previous limit for the relative entropy w.r.t
$\tilde\mu_\e$. Then, the result easily follows from the fact that
$\lim_{\e\to 0}\e^{-2}s(\mu_\ve|\tilde\mu_\e)=0$.

The method used in proving the hydrodynamic limit in the papers quoted
above is different from ours because it is based on the martingale
approach. However, in both methods it is needed an a priori estimate
of the entropy (which assures also the control of the Dirichlet form)
of the state of the process with respect to the invariant state of the
full dynamics. Since this is not known, one uses some trial reference
measure which is not invariant and as a result this entropy does not
decrease in time. 
The difficulty to get 
an entropy bound in the asymmetric case is
due to a diverging contribution to the flux of entropy due to the
asymmetric part of the generator. This flux has to be controlled by
the boundary generator. The choice of this generator is different from
the one used in the case of reversible bulk dynamics, where the
boundary death and birth process is required to satisfy a detailed
balance condition. Our generator is a generalization of the one used
by Derrida \cite{D}.  Our method would also provide the results for a
reversible boundary generator, but then we have to speed it up much
more than $\e^{-2}$.

One technical remark: the control of the terms in the bulk is done by
using the replacement Lemma and the non gradient result in
\cite{EMY1}. This requires a localization procedure for the currents,
which generates, in an open system, boundary terms. These terms are
dealt with, in a reversible non-gradient case \cite{LMS}, by a simple
use of integration by part and Schwartz inequality.  The version of
the integration by parts for the asymmetric case (Lemma 6.1 in
\cite{EMY1}) involves a function which is not bounded but just
summable and as a consequence we need rather to use a modified version
of the replacement Lemma to control these boundary terms.

We conclude this section by noticing that the extension to the
stationary problem of the convergence result in the macroscopic limit
could give some insight on the structure of the non-stationary states.
Recent papers \cite{DLS}, \cite{BDGJL}, focus on the problem of
characterizing the stationary non-equilibrium measures in terms of
large deviation functional.  Unfortunately, the relative entropy
method does not seem to be useful for the stationary problem, because
it relies on estimating the time derivative of the relative entropy in
terms of the entropy production which is expressed back again in terms
of the relative entropy, so that one gets a closed differential
inequality for the relative entropy. We think that the usual
martingale approach could be more suited for the stationary case. We
plan to refer on that in a future paper.

The plan of the paper is the following: in Section 2 we state the
results.  Section 3 is devoted to the strategy of the proof of the
entropy bounds and to the control of the bulk terms while in Section 4
we have collected the estimates of the boundary terms. Finally, in
Section 5 we discuss extensions to different geometries and different
boundary conditions.

\section{Notations and results}

Let $\ve>0$ such that $\ve^{-1}$ is integer and
$\L_\ve=\{-\ve^{-1},\ldots,\ve^{-1}\} \times \pi_\ve^{d-1}$ be the
cylinder in $\Z^d$, $d\ge 3$, of length $2\ve^{-1}+1$ with basis
$\pi_\ve^{d-1}$, the $(d-1)$-dimensional microscopic torus of size
$2\ve^{-1}+1$.  We denote by $\G_\ve=\{ x\in \L_\ve\,|\, x_1=\pm
\ve^{-1}\}$ the boundary of $\L_\ve$.  The elements of $\L_\ve$ will
be denoted by letters $x,y,\ldots$

A particle configuration is described as an element $\n\in
X_\ve=\{0,1\}^{\L_\ve}$, where $\n (x)=1, 0$ means that the site $x$ is
occupied by a particle or is empty.   
In this paper we are interested in the asymmetric simple exclusion process
(ASEP) on $\L_\ve$ with stochastic reservoirs at the boundary of $\L_\ve$. This
Markov process is defined through its infinitesimal generator 
\begin{equation}
  \label{eq:1}
\cl_\ve=\cl_{\ve,0}+\cl_{\ve,b}.
\end{equation}

The generator $\cl_{\ve,0}$ is  the nearest neighbor ASEP.
Its action on functions $f: X_\ve\to \R$ is 
\[
\cl_{\ve,0}f(\n)=\sum_{i=1}^d \sum_{x,x+e_i\in\L_\ve}r_{x,x+e_i} (\n) \left[
f(\n^{x,x+e_i})-f(\n)\right]
, \]
where $(e_1,\ldots,e_d)$ is the canonical basis of $\R^d$ and 
the rate functions $r_{x,x+e_i} (\n)$ are given by
$$
r_{x,x+e_i} (\n)=p_{e_i}\n(x)(1-\n(x+e_i)) +p_{-e_i}\n(x+e_i)(1-\n(x)) 
$$ 
and $\n^{x,y}$ is the 
configuration obtained
by exchanging the occupations of sites $x$ and $y$:
\[
\n^{x,y}(z)= \left\{ \begin{array}{lll}
 \n (y) & \text{if}\ z= x \,, \\
  \n(x) & \text{if}\ z=y \,, \\
   \n(z) & \text{if}\ z\not=x,y \, .\end{array}
     \right. 
\]
For $1\le i\le d$, $p_{e_i}$ ($p_{-e_i}$) is the jump probability of a
particle in the direction $e_i$ ($-e_i$). For convenience we normalize
the $p_e$'s so that, for $i=1,\ldots,d$, $p_{e_i}+p_{-e_i}=2$.  We
call $\d_i:=p_{e_i}-p_{-e_i},\ i=1,\ldots,d$, and we suppose that
$\d_1$ is non vanishing. We assume it positive without loss of
generality.

\smallskip
The currents $W_{x,i}$ are defined as 
\[
\begin{split}
W_{x,i}(\n) &=-W_{x,i}^{(s)} + W_{x,i}^{(a)}\\
\ & :=- \big[ \eta(x+e_i)-\eta(x) \big] + 
\d_i\left[\eta(x+e_i)\eta(x) -\frac{\eta(x) +\eta(x+e_i)}{2}   \right].
\end{split}
\]
\smallskip

Let $\L$ be the set $[-1,1 ]\times \pi^{d-1}$, where $\pi^{d-1}$ is
the $(d-1)$-dimensional torus with length 2, whose elements are denoted by
$u, v, \ldots$ We will consider as reference laws the Bernoulli 
product measures
$\nu_\rho$ on $X_\ve$ which are defined, for any smooth function $0<\rho<1$
on $\L$, by $\nu_\r(\n(x)=1)={\r(\ve x)}$.  It is well known that, in
infinite volume, the product measure of Bernoulli laws with any
constant parameter is invariant for the ASEP, but this is no longer valid
in finite volume.
\smallskip

The generator $\cl_{\ve,b}$ in (\ref{eq:1})  is the
infinitesimal generator of a birth and death process which creates
particles on the ``left'' of $\L_\ve$ and destroys them on the ``right''.
For any function $f$ on $X_\ve$
\[
\left(\cl_{\ve,b} f\right)(\n) =
    \sum_{x\in\G_\ve} C^b(\ve x,\eta) \big[f(\eta^x)
    -f(\eta)\big]\, ,
\]
where
$\n^{x}$ is the configuration obtained from
$\n$ by flipping the occupation number at site $x$
\[
\n^x(z)= \left\{ \begin{array}{ll}
 \n (z) & \text{if}\ z\ne x \\
  1-\n(x) & \text{if}\ z=x .\end{array}
     \right. 
\]
The rate functions are designed to fix the value of the particle
density at the boundary of $\L_\ve$.  Let $\Gamma_{\ve,+}$ (resp.
$\Gamma_{\ve,-}$) be the ``left'' (resp. ``right'') side of
$\Gamma_\ve$:
\[
\begin{split}
\Gamma_{\ve,+} &
=\left\{ (-\ve^{-1},x_2,\dots,x_d)\ ,\ \ (x_2,\dots,x_d)\in
\pi_\ve^{d-1}\right\}, \\
\Gamma_{\ve,-} &=\left\{ (\ve^{-1},x_2,\dots,x_d)\ ,\ \ (x_2,\dots,x_d)\in
\pi_\ve^{d-1}\right\}\, .
\end{split}
\]
For a smooth function $b$ on the boundary $\Gamma$ of $\L$ that we fix
from now on, we choose $C^b(\ve x,\eta)$ such that a particle is added
on $x\in\Gamma_{\ve,+}$, when the site is empty, with rate $\d_1(1/2
+\ve b(\ve x))$ and removed in $\Gamma_{\ve,-}$, when the site is
occupied, with rate $\d_1(1/2 -\ve b(\ve x))$:
\[
C^b (\ve x, \n)      = \d_1
\bigl(1/2+\ve b (\ve x)\bigr)(1-\n (x))\1\{x\in\Gamma_{\ve,+}\} +\d_1
\bigl(1/2-\ve b (\ve x) \bigr) \n (x) \1\{x\in\Gamma_{\ve,-}\}\, .
\]
To understand this choice, remark that if we take $b=0$
then the density $1/2$ is stationary
for the full dynamics $\cl_\ve$, therefore the leading coefficient $1/2$ in
the rate is necessary for the incompressible limit to make sense. The
term $\ve b$ will fix the value at the boundary of the density perturbation
with respect to the stationary value $1/2$.

\smallskip
We will study the dynamics defined by the generator $\cl_\ve$
under the diffusive space-time scaling $(\ve x,\ve^{-2}t)$. We denote by
$\n_t$ the particle configuration at time $t$ of the associated Markov process.
When the process starts from the product measure 
$\mu_\ve=\nu_{1/2+\ve m(.)}$,
$m$ a smooth function on $\L$ such that $m{\big|_\G}=b$,
its law is denoted by $\mathbb{P}_\ve^{b,m}$.

As in [Y] and [EMY1], the proof is based on the study of the relative entropy.
Given two measures $\mu_1$ and $\mu_2$ on $X_\ve$ (with $\mu_1$ absolutely
continuous w.r.t. $\mu_2$),
the entropy of $\mu_1$ with respect to $\mu_2$ is
defined as
\[
H(\mu_1|\mu_2)=\int\frac{d\mu_1}{d\mu_2}\log\frac{d\mu_1}{d\mu_2}\, d\mu_2\, .
\]
We also introduce the relative entropy $s(\mu_1|\mu_2):=\ve^d
H(\mu_1|\mu_2)$.

For simplicity we call $\nu:=\nu_{1/2}$ the reference measure.  
Notice that the measure $\mu_\ve$ has a density $\Psi$ with respect to
$\nu$ given by
$$
\Psi(\n)=\frac{1}{Z}\exp\left\{ \sum_{x\in\L_\ve} \ve
  \varphi(m(\ve x)) \n (x)\right\} \nu(\n),
$$
where 
\begin{equation}
\label{eq:varphi}
\varphi(u) =\ve^{-1} \log\left\{ \frac{1+2\ve m(u)}{1-2\ve m(u)}
\right\} 
\end{equation}
and $Z$ is a normalization constant.
Then the relative entropy of the measure $f\nu$ (where $f$ is some
probability density) w.r.t. $\mu_\ve$ can be written as:
\[
s(f|\Psi):=s(f\nu|\mu_\ve) =\ve^d \int f \log\frac{f}{\Psi}\, d\nu .
\] 
The probability $\mu_\ve (t)$ is the law of $\n_t$ when the initial
distribution of the process is $\mu_\ve$. Its density with respect 
to $\nu$ is denoted by $f_t$.
We now consider the solution $m(t,u)$ of the following partial
differential equation 
\begin{equation}
\label{burgbound}
  \left\{ \begin{array}{lll}
 \partial_t m (t,u) & =\sum_{i=1}^d \delta_i \partial_{u_i}\big( m(t,u)\big)^2
+\sum_{1\le i,j\le d}D_{i,j}\partial^2_{u_i,u_j}m(t,u) \\
  m(0,.) & = m \\
   m(t,.){\Big|_\G} & =b(.) \quad\text{for}\ t\ge 0 ,\end{array}
     \right.\qquad\qquad 
\end{equation}
where the diffusion coefficients $D_{i,j}$ will be defined later on. 

We introduce the measure $\nu_{1/2+\ve m(t,.)}$ whose density
w.r.t. $\nu$ is given by
\[
\Psi_t(\n) =\frac{1}{Z_t}\exp\left\{ \sum_{x\in\L_\ve} \ve
  \varphi(m(t,\ve x)) \n (x)\right\},
\]
with $Z_t$ a normalization constant.

\smallskip
Our main result is:
\begin{thm}
\label{th1} Let $s(t)=s(f_t| \Psi_t)$. Then,
for any $t>0$,
\[
\lim_{\ve \rightarrow 0} \ve^{-2} s(t) =0\, .
\]
\end{thm}

Using standard arguments based on the entropy inequality (see e.g. [Y]
and [EMY1]), we get as a corollary of the previous theorem the
incompressible limit of ASEP with stochastic reservoirs:

\begin{thm}
Let $\pi_t^{\ve}$ be the empirical measure 
\[
\pi_t^{\ve}(\n) =\frac{1}{|\L_{\ve}|}\sum_{x\in\L_{\ve}}
\Bigl(\n_t(x) -\frac{1}{2}\Bigr)
\delta(\ve x).
\]
Then $\pi_t^{\ve}$ converges weakly in
$\mathbb{P}_\ve^{b,m}$-probability to the solution $m(t,u)$
to (\ref{burgbound}).
\end{thm}

\bigskip
\section{Bounds on the entropy }
The strategy in proving Theorem \ref{th1}. is based on the study of the
time evolution of the relative entropy. 
Here the difficulty relies on the fact that the invariant
measure for the full dynamics is unknown. 

\medskip Actually, we will not work directly with the entropy $s(t)$
but with some approximation of $s(t)$ obtained by a modification of
the density $\Psi_t$.  Let $m$ be the solution of the equation
(\ref{burgbound}) and $\g$ be a smooth function defined on a
neighborhood of $\L$ such that $\partial_{u_1}\gamma$ has a compact
support included in $]-1,1[\times \pi^{(d-1)}$ with
$\g(.){\big|_\G}=b$.  We introduce a new parameter $0< \t <1 $ which
will go to 0 after $\ve$, we denote by $\Omega_\t$ the set
\begin{equation}
  \label{eq:7}
\Omega_\t:=[-1+\t,1-\t]\times \pi^{(d-1)}  
\end{equation}
and we choose a smooth function
$\chi^\t$ in a neighborhood of $\L$ such that
$\chi^\t{\Big|_{\Omega_{2\t}}}=1$ and $\chi^\t=0$ outside $\Omega_\t$.
Then we define
\[
\rho(t,u)=m(t,u) \chi^\t(u) +\gamma(u) \bigl( 1-\chi^\t(u) \bigr).
\] 
Notice that $\rho(t,.)$ is identically equal to $m(t,.)$ in
$\Omega_{2\t}$ and equal to $\gamma$ on $\Omega\setminus \Omega_\t$.

Recalling that $\varphi$ was defined in (\ref{eq:varphi}), we
denote by $\l(t,u)$ the function
$$
\l(t,u)=\varphi\bigl(\rho(t,u)\bigr),
$$
then it is easy to check that, for $u\in\Omega_{2\t}$,
\begin{equation}
  \label{eq:8}
   \partial_t \l(t,u) -\sum_{i=1}^d \delta_i \partial_{u_i}\big( \l(t,u)\big)^2
-\sum_{1\le i,j\le d}D_{i,j}\partial^2_{u_i,u_j}\l(t,u)=o(\ve),
\end{equation}
where $\ve^{-1}o(\ve)$ is a bounded function.
Moreover, with a suitable choice of $\chi^\t$, we may assume that
the derivative $\partial_{u_i}\l$ , $i=1,\ldots,d$ are bounded,
uniformly in $\t$.

Following \cite{EMY1}, we modify suitably the density $\Psi_t$ in the
definition of the relative entropy $s(t)$: for an integer $\ell$, we
set $k=\ell \ve^{-2/d}$ and $\L_k=\{-k,\cdots,k \}^d$.  
The normalized indicator function $\omega$ is defined as
$$
\omega(x)=|{\L}_k|^{-1} \1\{x\in {\L}_{k} \}.
$$

Let $\{F_n\}$ be a family of local functions (i.e. which depend on the
particle configuration only through a finite number of sites) on $X_\ve$  and 
$$ 
\Phi_n(\n) =\sum_{i=1}^d \sum_{x\in \L_\ve^{2k}} \big( \partial_{u_i}
\l \big) (t,\ve x) \big( \tau_x F_n* {\omega}\big),
$$
for $x\in \Z^d$, where $\tau_{\,\cdot\,} $ is the shift operator
on $\X=\{ 0,1\}^{\Z^d} $,
$\L_\ve^{2k}=\{-\ve^{-1}+2k,\ve^{-1}-2k \}\times \pi^{d-1}$
and $*$ is the convolution
product. We will work with $n$ fixed and will take the
limit $n\to \infty$ after the limit $\ve\to 0$.
Therefore the index $n$ will be omitted for sake of shortness.
The modified density $\wt{\Psi}_t^F$ is defined as in \cite{EMY1} by 
\[
\wt{\Psi}_t^F(\n)= \frac{1}{\wt{Z}_t^F}\exp\left\{ \sum_{x\in\L_\ve} \ve
  \left( \l *\omega\right)(t,\ve x) \n (x) +\ve^2 \Phi(\eta)\right\},
\]
where $\wt{Z}_t^F$ is the normalization constant.

Moreover, we define 
\[
s_1(t):=s(f_t|\wt{\Psi}_t^F),\ \ 
 h_t^F=\frac{f_t}{\wt{\Psi}_t^F},\ \  h_t=\frac{f_t}{\wt{\Psi}_t},
\]
and $\wt{\Psi}_t$ the density 
$$
\wt{\Psi}_t(\n)= \frac{1}{\wt{Z}_t}\exp\left\{ \sum_{x\in\L_\ve} \ve
  \left( \l *\omega\right)(t,\ve x) \n (x) \right\}.
$$

We finally introduce 
the functionals $D_0$, $D_b$ which are
closely related to the Dirichlet forms. For
any non negative function $h$ and any measure $\mu$ on $X_\ve$,
\begin{equation}
\begin{split}  
D_0\big(h, \mu\big) & =
\sum_{i=1}^d\sum_{x,x+e_i\in\L_\ve } \int
     r_{x,x+e_i}(\n)
  \left( {h}(\n^{x,x+e_i}) -{h}(\n) \right)^2 d \mu \, ,\\
D_b\big(h,\mu) & =\frac{\d_1}{2}
\sum_{x\in\Gamma_\ve} \int  
  \left( {h}(\n^{x}) -{h}(\n) \right)^2 d \mu \, .
\end{split}
\end{equation}

Then we have


\begin{prop}\label{eq:prop2} 
There exists some constant
$K>0$ such that for every $t>0$,
$$
\lim_{\t\to 0}\limsup_{n\to\infty}
\limsup_{\ell \rightarrow \infty}\limsup_{\ve \rightarrow 0} \left\{
\ve^{-2} s_1(t) +K\,\ve^{d-4} \int_0^t
  ds \left(
     D_0\big(\sqrt{h_s^F}, \wt{\Psi}_s^F\nu\big)+
     D_b\big(\sqrt{h_s},\wt{\Psi}_s \nu\big) \right)
    \right\}=0.
$$
\end{prop} 

\medskip
\noindent{\bf Remark.}
Since the Dirichlet forms $D_0$ and $D_b$ are positive, Theorem
\ref{th1} will follow from Proposition \ref{eq:prop2} if we can
show that 
\begin{equation}
\label{eq:2}
\limsup_{\t\to 0}\limsup_{\ell \rightarrow \infty}\limsup_{\ve \rightarrow
  0}\ve^{-2}\bl[s(t)-s_1(t)\br]\le 0.
\end{equation}
Consider the density $\Psi^1_t$ with respect to $\nu$ given by
\begin{equation}
  \label{eq:2.1}
\Psi^1_t(\n)= \frac{1}{Z^1_t}\exp\left\{ \sum_{x\in\L_\ve} \ve
  \l(t,\ve x)\n(x)\right\}.  
\end{equation}
with $Z^1_t$ the normalization constant.
It is proved in Lemma 3.2 of \cite{EMY1} that for any $0<\t<1$
\begin{equation}
  \label{eq:3}
\lim_{\ve \rightarrow 0}\ve^{-2}\bl[s\bl(f_t|\Psi^1_t\br)-s_1(t)\br]=0.
\end{equation}
Moreover a straightforward generalization of the proof of this lemma
together with the entropy inequality leads to the following estimate:
for any $\a>0$, there exists a constant $c(\a)$ such that
\begin{equation}
  \label{eq:4}
\ve^{-2}\bl[s(t)-s\bl(f_t|\Psi^1_t\br)\br]\le
\a\ve^{-2}s\bl(f_t|\Psi^1_t\br)+c(\a)\t  
\end{equation}
where we have used the fact that $\l(t,u)=\vp(m(t,u))$ for any $t>0$
and $u\in \Omega_{2\t}$. Then (\ref{eq:3}) and (\ref{eq:4}) imply (\ref{eq:2}).

\bigskip
\noindent{\bf Proof of Proposition \ref{eq:prop2}.}
A simple computation shows that
\begin{equation*}
  \begin{split}
    \ve^{-2}\frac{d}{dt}s_1(t)  & =
    \ve^{d-4}\int f_t \cl_\ve\log h_t^F d\nu
    -\ve^{d-2}\int \frac{\partial_t
    \wt{\Psi}_t^F}{\wt{\Psi}_t^F} f_t d\nu\\  
  \end{split}
\end{equation*}

Since the function $\Phi(\eta)$ depends on the configuration $\eta$
only through the variables $\left\{\n(x_1,\cdots,x_d)\ ,\  |x_1|\le (\ve^{-1}- k)
  +s_F  \right\}$, with $s_F$ the support of $F$, 
then
\[
\ve^{d-4}\int f_t \cl_\ve \log h_t^F d\nu=
\ve^{d-4}\int f_t \cl_{\ve,0} \log h_t^F d\nu
+ \ve^{d-4}\int f_t
\cl_{\ve,b}   \log h_t  d\nu .
\]

We use the basic inequality
\[
a\log\frac{a}{b}\le -\bigl(\sqrt{a}-\sqrt{b}\bigr)^2+a-b
\]
for $a$ and $b$ positive numbers. It is applied with $a=h^F_t(\eta^{x,x+e_i}) $
and $b=h^F_t(\eta)$ to manage with $\cl_{\ve,0}$ and with
$a=h_t(\eta^{x})$ and $b= h_t(\eta)$ to manage with $\cl_{\ve,0}$, We
get
\begin{equation}
 \label{eq:entr1}
\begin{split}
  \ve^{-2}\frac{d}{dt}s_1(t) &\le  E(\ve,t) 
       -\ve^{d-4}D_0\big(\sqrt{h_t^F}, \wt{\Psi}_t^F\nu) \\
\ &\ \ +  \ve^{d-4}\int \big( \cl_{\ve,b}  h_t \big) \wt{\Psi}_t    d\nu
   -\frac{1}{2}\ve^{d-4}  D_b(\sqrt{h_t},\wt{\Psi}_t\nu)  
\end{split}
\end{equation}
with $E(\ve,t):= E_1+E_2$,
where
\begin{equation*}
  \begin{split}
    E_1 &:= \ve^{d-4} \int \frac{\cl_{\ve,0}^*\wt{\Psi}_t^F
      }{\wt{\Psi}_t^F} f_t d\nu = \ve^{d-4} \int
    \frac{\cl_{\ve,0}^*\wt{\Psi}_t^F }{\wt{\Psi}_t^F} 
    \big( f_t-\wt{\Psi}_t^F  \big) d\nu\\
    E_2 &:= -\ve^{d-2}\int
    \frac{\partial_t\wt{\Psi}_t^F}{\wt{\Psi}_t^F} f_t d\nu
 \end{split}
\end{equation*}
and $\cl_{\ve,0}^*$ denotes the adjoint operator of
$\cl_{\ve,0}$ in $L^2\big( \nu\big)$. 
After a change of variables the term $E_1$ can be rewritten as
\begin{equation}
  \begin{split}
E_1 & =\ve^{d-4} \sum_{i=1}^d \sum_{x,x+e_i\in\L_\ve}\int r_{x,x+e_i}^* (\n) 
\left\{
  \frac{\wt{\Psi}_t^F (\n^{x,x+e_i})}{\wt{\Psi}_t^F (\n)} -1
  \right\} \big( f_t - \wt{\Psi}_t^F \big)(\n) d \nu\\
\ &\ \  + \ve^{d-4}\sum_{i=1}^d \sum_{x,x+e_i\in\L\ve} \d_i \int
  \big(\n(x+e_i)-\n(x) \big) \big( f_t -\wt{\Psi}_t^F \big) (\n) d \nu\\
\ &:= E_1^1 +E_1^2,
\end{split}
\end{equation}
where 
$$
r_{x,x+e_i}^*(\n) 
    = p_{e_i} \n(x+e_i) \big( 1-\n(x) \big)
                  +p_{-e_i} \n(x) \big( 1-\n(x+e_i) \big).
$$
Using again a change of variables at the boundary,
the second term of $E_1$ is equal to 
\begin{equation}
  \begin{split}
E_1^2& =\ve^{d-4}\sum_{x\in\Gamma_\ve} \d_1 \int
  \big(\n(x)- \big<\eta(x) \big>_{\wt{\Psi}_t^F} \big) 
         {n}_1 (\ve x) f_t(\n)d \nu\\
& =\ve^{d-4}\sum_{x\in\Gamma_\ve} \d_1 \int
  \big(\n(x)- \big<\eta(x) \big>_{\wt{\Psi}_t} \big) 
         {n}_1 (\ve x) f_t(\n)d \nu\, ,
\end{split}
\end{equation}
where $n=(n_1,0,\cdots ,0)$ is the
outward unit normal vector to the boundary surface $\Gamma$, and 
by standard manipulations can
be bounded by the Dirichlet form and the generator at the boundary:

\begin{lem}
\label{eq:lembord}
For all $a>0$,
\[
\begin{split}
& \ve^{d-4}\sum_{x\in\Gamma_\ve} \d_1 \int
  \big(\n(x)- \big<\eta(x) \big>_{\wt{\Psi}_t} \big) 
         n_1 (\ve x) f_t(\n)d \nu \,\le \\
 &\qquad\qquad\qquad
   \le -\ve^{d-4}  \int \big( \cl_{\ve,b} h_t \big)\wt{\Psi}_t d \nu 
+\ve^{d-4}\frac{a}{2} D_b(\sqrt{h_t},\wt{\Psi}_t\nu)+\ve^{3}k^4
  \frac{C_1}{a}\, ,
\end{split}
\]
for some positive constant $C_1$.
\end{lem}

\begin{proof}
For $\ve >0$, denote by ${\wt{\cl}}_{\ve,b}^k$ the following
generator at the boundary
\[
\begin{split}
\Big( {\wt{\cl}}_{\ve,b}^k f\Big) (\n) &=
 \d_1\sum_{x\in\Gamma_{\ve,+}}{\wt{C}}_+^b(\ve x,\n)
  \big[ f(\n^{x})-f(\n) \big]\\
\ & \ \ +\d_1\sum_{x\in\Gamma_{\ve,-}}{\wt{C}}_-^b(\ve x,\n)
  \big[ f(\n^{x})-f(\n) \big]\, ,
\end{split}
\]
where
\[
\begin{split}
{\wt{C}}_+^b(\ve x,\n) & =
     \left(1/2+\ve \big(\l *\omega \big) (t,\ve x)\right)(1-\n (x))\\
{\wt{C}}_-^b(\ve x,\n) & =
     \left(1/2-\ve \big(\l *\omega \big) (t,\ve x)\right)\n (x)
\end{split}
\]

Remark that the replacement of $\cl_{\ve,b}$ by  ${\wt{\cl}}_{\ve,b}^k$
in the derivative of the entropy produces a term that we can bound
by the Dirichlet form. Indeed, we have
\[
\begin{split}
&\ve^{d-4}\int \left(\Big( \cl_{\ve,b}- {\wt{\cl}}_{\ve,b}^k\Big)h_t
    \right) \wt{\Psi}_t d\nu =\\
 &\qquad\qquad
   =\ve^{d-3}   \sum_{x\in\Gamma_{\ve,+}} \Big( b(\ve x)- (\l *\omega)(t,\ve)\Big)\int
   (1-\eta(x))\big[ h_t(\n^x)-h_t(\n)\big]\wt{\Psi}_t d \nu\\
&\qquad\qquad
\ \ + \ve^{d-3}   \sum_{x\in\Gamma_{\ve,-}} \Big( (\l *\omega)(t,\ve)
     -b(\ve x)\Big)
  \int \eta(x)\big[ h_t(\n^x)-h_t(\n)\big]\wt{\Psi}_t d \nu\, ,
\end{split}
\]
by Taylor expansion at the second order, this last quantity is equal to
\[
\begin{split}
&\ve^{d-1}k^2   \sum_{x\in\Gamma_{\ve,+}} 
G(t,\ve x) \int  (1-\eta(x))\big[ h_t(\n^x)-h_t(\n)\big]\wt{\Psi}_t d \nu\\
&\ \ \ + \ve^{d-1}k^2 \sum_{x\in\Gamma_{\ve,-}} G(t,\ve x)
  \int \eta(x)\big[ h_t(\n^x)-h_t(\n)\big] \wt{\Psi}_t d \nu\\
&\le \ve^{d-4}\frac{a}{2}D_b(\sqrt{h_t},\wt{\Psi}_t\nu)+\ve^{3}k^4
  \frac{C_1}{a}
\end{split}
\]
for all $a>0$ and some constant $C_1$, where $G(t,\ve x)$ is bounded
and proportional to the second derivative of $\l(t,.)$. We have
used, in the last inequality, Schwartz inequality and the fact that
$h_t$ is a probability density with respect to $\wt{\Psi}_t\nu$.

To conclude the proof of the lemma, it is easy to check that
the left hand side of the inequality in Lemma
\ref{eq:lembord} can be rewritten as
$-\ve^{d-4}\int \left({\wt{\cl}}_{\ve,b}^k h_t
    \right) \wt{\Psi}_t d\nu\, $, by using the change of variable formulas valid for any measure $\nu_g$ 
\[
  \begin{split}
     \int\bigl(1-\n_x\bigr)f(\n^{x,+})\,d\nu_g(\n)
    &=\frac{1-g(\ve x)}{g(\ve x)}\int\n(x)f(\n)\,d\nu_g(\n) \\
     \int\n_xf(\n^{x,-})\,d\nu_g(\n)
    &=\frac{g(\ve x)}{1-g(\ve x)}\int\bigl(1-\n(x)\bigr)f_\n\,d\nu_g(\n)  
  \end{split}
\]
where $\n^{x,+}$ and $\n^{x,-}$ are the configurations obtained from
$\n$ by adding or removing a particle at the site $x$. 
\end{proof}


The term $E_1^1$ is shown, 
by using
Taylor expansion and Schwarz inequality, to be less or equal to
\begin{equation}
\label{eq:E11}
 \ve^{d-2} \int \left\{ \sum_{i=1}^d \sum_{x\in\L_\ve} 
\big( \partial^{\ve,i}\l*\omega \big) (t,\ve x)
 \wh{W}_{x,i} -\cl_0^* \Phi  \right\}
\big( f_t - \wt{\Psi}_t^F \big)(\n) d \nu
+R_\ve(t) \, + o_\ve (1).
\end{equation}
where
$W_{x,i}$ is the current defined in section 2, $\partial^{\ve,i}\l$ is the
discrete gradient of $\l$
$$
\partial^{\ve,i}\l(t,u)=\ve^{-1}\Big[\l(t,u+\ve e_i)-\l (t,u)\Big]
$$
and for any function $g$ on $X_\ve$, we
denote by $\wh{g}=g-<g>_\nu $.
The operator
$\cl_0$ is the generator of the ASEP in infinite volume space $X=
\Z\times \pi_\ve^{d-1}$ and $\cl_0^*$ is its adjoint in $L^2(\nu)$,
given for any local function $f$ by 
$$
\big(\cl_0^* f\big)(\eta)=\sum_{i=1}^d
\sum_{x\in \Z\times \pi_\ve^{d-1}}
r_{x,x+e_i}^*(\eta)\left[f(\n^{x,x+e_i}) -f(\n) \right].
$$
In equation \eqref{eq:E11} as in the sequel, $R_\ve(t)$ is
real sequence of the form 
\begin{equation}
\label{eq:rest}
R_\ve(t)=
\ve^{d} \sum_{x\in\L\ve} G (t,\ve x)
\int \tau_x \wh{ g}(\n) 
\big( f_t - \wt{\Psi}_t^F \big)(\n)  d \nu 
\end{equation}
with some bounded functions $G(t,.)$ defined on $\L$ and $g$ on $X_\ve$.
Observe  that, for such sequences, one can use the entropy inequality and obtain,
for any $A>0$,
\begin{equation}
\label{eq:rest-entr}
R_\ve(t)\le A \ve^{d-2} s_1(t) + A^{-1}o_\ve(1)\, ,
\end{equation}
for a real sequence $o_\ve(1)$ which is bounded in absolute value by
a constant that converges to 0 as $\ve\downarrow 0$.

At this point, we would like to replace
the currents $W_{x,i}$ appearing in the first term of the
right hand side of (\ref{eq:E11})  by its convolution with $\omega$. 
In order to do it we need some notation. For a local function $g$ denote by
$g^k$ the convolution 
$$
\big(\tau_x g\big)^k (\n)=\frac{1}{|{\L}_k|}\sum_{y\in{\L}_k}
  \big(\tau_{x+y} \wh{g}\big)(\n)\, ,
$$
when $g(\n)=\n(0)-1/2$, we shall denote $\big(\tau_x g\big)^k $ simply by
$\eta^k (x)$. Observe that, $\eta^k(0) +1/2$ is precisely the empirical density in 
${\L}_k$. Furthermore, 
for $\ve >0$, 
let $\L_\ve^{k}=\{-\ve^{-1}+k,\ve^{-1}-k \}\times \pi^{d-1}$ and
$\Gamma_\ve^k=\L_\ve \setminus \L_\ve^{k}$.
For $x\in X=\Z\times \pi^{d-1}$ denote by $\overline{W}_{x,i}$ the current
defined on $\X=\{0,1\}^X$ by
\[
\overline{W}_{x,i}(\n)= \left\{ \begin{array}{ll}
W_{x,i} (\n) & \text{if}\ x,x+e_i\in \L_\ve  \\
  0 & \text{if}\  x\ \text{or}\ x+e_i\notin \L_\ve \end{array}
     \right. 
\]
with this notation we have
\begin{equation}
\label{eq:localisation}
\begin{split}
E_1^1 &= \ve^{d-2} \int \left\{ \sum_{i=1}^d \sum_{x\in\L_\ve^k}
  \partial^{\ve,i} \l(t,\ve x)
 \big(\tau_x\wh{W}_{0,i}\big)^k -\cl_0^* \Phi  \right\}
\big( f_t - \wt{\Psi}_t^F \big)(\n) d \nu\\
&\ \ + \ve^{d-2} \int \left\{ \sum_{i=1}^d \sum_{x\in\Gamma_\ve^k} 
\partial^{\ve,i}\l(t,\ve x)
 \big(\wh{\overline{W}}_{x,i}\big)^k   \right\}
\big( f_t - \wt{\Psi}_t^F \big)(\n) d \nu\\
\ & \ :=E_1^3+E_1^4\, .
\end{split}
\end{equation}

Let us summarize what we have done so far: for
any fixed $A>0$ and $a>0$, we got the following bound for the entropy derivative
\begin{equation}
\label{eq:3.9}
\begin{split}
\frac{d}{dt} \ve^{d-2} s_1(t) & \le A \ve^{d-2} s_1(t) +
E_2 + E_1^3 + E_1^4\\
& \ \ - \ve^{d-4}\left\{ D_0\big(\sqrt{h_t^F}, \wt{\Psi}_t^F\nu_{1/2}\big)
    +\frac{1-a}{2} D_b\big(\sqrt{h_t},\wt{\Psi}_t \nu_{1/2}\big) \right\}\\
& \ \ +o_\ve(1)\, .
\end{split}
\end{equation}

>From the fact that $\partial^{\ve,1} \l(t,\ve x)$ has a compact support,
Taylor expansion and (\ref{eq:rest-entr}), the  term $E_1^3$ can be written as
\begin{equation}
\begin{split}
E_1^3 &= \ve^{d-1} \sum_{i=1}^d \sum_{x\in\L_\ve^k} \partial_{u_i}^2
        \l (t,\ve x)
        \int \eta^k(x)
     \big( f_t - \wt{\Psi}_t^F \big) (\n)  d \nu \\
\ &\ \  +\ve^{d-2} \int \left\{ \sum_{i=1}^d\sum_{x\in\L_\ve^k}
  \partial^{\ve,i} \l(t,\ve x)
         \big({\wh{W}}_{x,i}^{(a)}\big)^k -\cl_0^* \Phi  \right\}   \big( f_t -
        \wt{\Psi}_t^F \big) (\n)  d \nu \\
\ & \ \ +o_\ve(1)\, .
\end{split}
\end{equation}

\smallskip 
Following the method of \cite{EMY1} we now replace the currents
$\big({\wh{W}}_{0,i}^{(a)}\big)^{k}$ in the bulk by a linear
combination of the gradients $\big( \n(e_j)-\eta(0)\big)^k$,
$j=1,\cdots ,d$. This requires some notation.  For $x\in\L_\ve^k$ and
$1\le i\le d$, denote by $M_{x,i}^k$ the conditional expectation of
$W_{x,i}^{(a)}$ given the density of particles on
$\L_{x,k}$:
$$
M_{x,i}^k(\n)=
\E \left[  W_{x,i}^{(a)} \big| \eta^k(x)\right]\, ,
$$
where ${\L}_{x,k}=\{x+y\  :\   y\in  {\L}_k\}$.
One can compute $M_{x,i}^k(\n)$ easily. It is given by
$$
\d_i\Big( 1+ \frac{1}{(2k+1)^d-1}\Big)
\n^k (x)\big(\n^k(x)-1  \big)\, .
$$

Furthermore, for $\ve >0$, $1\le i\le d$  and $\n\in X_\ve$, let
$$
\V_i^k(\n) =\Big( \big(\wh{W}_{0,i}^{(a)}\big)^k-M_{0,i}^k\Big) (\n) 
  -\big(\cl_0^* F\big)^k
     +\sum_{j=1}^d \wt{D}_{i,j}\times  \left( \n^k(e_j)-\n^k(0)\right), 
$$
where $\wt{D}_{i,j}=D_{i,j}-\d_{i,j}$ and the diffusion matrix
$D_{i,j}$ is defined in Section 2.

\medskip
The next result is the main step towards the proof of the bounds of
the entropy.

\begin{thm} 
\label{eq:EMYa}
For all $A>0$ and any probability density $f$ with
respect to $\nu$,
$$
\limsup_{n\to \infty}\limsup_{\ell \to \infty} \limsup_{\ve\to 0}
\left\{
\ve^{d-2}\int  \Big( \sum_{i=1}^d\sum_{x\in \L_\ve^k} \partial^{\ve,i}\l(t,\ve x)
\tau_x \V_i^k (\n) \Big)f(\eta) d \nu -A \ve^{d-4} D_0(\sqrt{f},\nu) 
        \right\} \le 0\, .
$$
\end{thm}
\bigskip

This Theorem was proved in \cite{EMY1}, [LY]. The next lemma takes care of the 
terms close to the boundary ($E_1^4$)
\begin{lem}
\label{eq:rest-local}
For all $A>0$ and any probability density $f$ with
respect to $\nu$,
\[
\limsup_{\ell\to \infty}\limsup_{\ve \to 0}\Bigg\{
\ve^{d-2} \int \sum_{i=1}^d \sum_{x\in\Gamma_\ve^k} 
\partial^{\ve,i} \l(t,\ve x)
\big(\wh{\overline{W}}_{x,i}\big)^k
 f\, d \nu
 -A \ve^{d-4} D_0(\sqrt{f},\nu)  \Bigg\}
   \le 0\, .
\]
\end{lem}
The proof of this Lemma will be given in next section.
\medskip

Notice that, in the previous results, the Dirichlet form
$D_0(\sqrt{f_t},\nu)$ appeared while, in the derivative of the
entropy, we got $D_0(\sqrt{h_t^F},\wt{\Psi}_t^F \nu)$.
The next lemma allows us to replace the one by the other. Moreover we
will also need some estimate about $D_0(\sqrt{\wt{\Psi}_t^F}, \nu)$.

\begin{lem}\label{eq:Dirichlet}
There exist two  constants $C_0$ and $C_0'$ such that
\[
\begin{split}
-\ve^{d-4}D_0\big(\sqrt{h_t^F},\wt{\Psi}_t^F \nu \big)  &\le -\frac{1}{2}\ve^{d-4} D_0(\sqrt{f_t},\nu)
 +C_0\, , \\
\ve^{d-4}D_0\big(\sqrt{\wt{\Psi}_t^F}, \nu \big)  &\le C_0'
\end{split}
\]
\end{lem}

\begin{proof}
The second inequality follows by inspection.
For two nearest neighbor sites $x,y$ in $\L_\ve$, denote by $S_{x,y}$
the operator defined by
$$
\big(S_{s,y}f\big) (\n)= f(\n^{x,y})-f(\n).
$$
and write the Dirichlet form $D_0(\sqrt{f_t},\nu)$ as
\[
\begin{split}
D_0\big(\sqrt{f_t},\nu \big) =
\sum_{i=1}^d \sum_{x,x+e_i\in\L_\ve} \int r_{x,x+e_i} 
&\Big\{ \Big( S_{x,x+e_i}  \big(\wt{\Psi}_t^F\big)^{-\frac{1}{2}}\Big)(\n) 
      \sqrt{f_t}(\n^{x,x+e_i}) \\
\ &
+  \Big(S_{x,x+e_i} \sqrt{h_t^F}\Big)(\n) \Big\}^2
\wt{\Psi}_t^F(\n) d\nu .
\end{split}
\]

The elementary inequality $(a+b)^2\le 2\big( a^2+b^2\big)$ and a
change of variables give that
\[
\begin{split}
\ve^{d-4}D_0\big(\sqrt{f_t},\nu \big) & \le
  2\ve^{d-4} D_0\big(\sqrt{h_t^F},\wt{\Psi}_t^F \nu \big) \\
\ & \ \ +2\ve^{d-4} \sum_{i=1}^d\sum_{x,x+e_i\in\L_\ve}
\int r_{x,x+e_i}^*(\n)
  \left(\sqrt{\frac{\wt{\Psi}_t^F(\n^{x,x+e_i})}{\wt{\Psi}_t^F(\n)}}-1
  \right)^2 f_t(\n) d\nu \, .\\
\end{split}  
\]
\end{proof}

Applying Lemma \eqref{eq:rest-local} with densities $f_t$ then
  $\wt{\Psi}_t^F$ and using Lemma \ref{eq:Dirichlet}, we obtain
the following bound for $E_1^4$: for any $A>0$
\begin{equation}
  \label{eq:10}
  E_1^4\le A\ve^{d-4} D_0(\sqrt{f_t},\nu)+AC_0'+r(\ve,k,A)
\end{equation}
where 
\[
\limsup_{\ell\to\infty}\limsup_{\ve\to 0}r(\ve,k,A)\le 0.
\]

With the notations introduced before Theorem \ref{eq:EMYa},
a summation by parts and a Taylor expansion permit to rewrite the quantity 
$E_1^3$ as
\begin{equation}
  \label{eq:6}
\begin{split}
E_1^3  &=
  \ve^{d-1} \int \Big\{  \sum_{1\le i,j\le d} \sum_{x\in \L_\ve^k}
          \partial_{u_i,u_j}^2 \lambda (t,\ve x) D_{i,j} \n^k(x)
          \Big\} \left( f_t -\wt{\Psi}_t^F  \right) (\n) d\nu \\
 & \ \ +\ve^{d-2}\int \Big\{
\sum_{i=1}^d\sum_{x\in\L_\ve^k} \partial^{\ve,i} \l(t,\ve x)
M_{x,i}^k(\n)\Big\}
\left( f_t - \wt{\Psi}_t^F  \right) (\n)
          d \nu  \\
& \  \ +\ve^{d-2} \int  \Big\{ \sum_{i=1}^d\sum_{x\in \L_\ve^k} 
\partial^{\ve,i}\l(t,\ve x)
\tau_x \V_i^\ve (\n) \Big\} \left( f_t - \wt{\Psi}_t^F  \right) (\n)
          d \nu \\
&\ \ +\ve^{d-2}\int \Big\{\sum_{i=1}^d  \sum_{x\in\partial\L_\ve^k}
     \partial^{\ve,i} \l(t,\ve x)(t,\ve x) \n^k(x)
    n_1(\ve x) \Big\}\left( f_t - \wt{\Psi}_t^F  \right) (\n)
          d \nu  \\
& \  \ + R_\ve (t) + o_\ve(1)\, ,
\end{split}  
\end{equation}
where $\partial\L_\ve^k$ stands for the boundary of $\L_\ve^k$,
$n(\ve x)=(n_1(\ve x),0,\cdots,0)$ for 
the outward unit normal vector to $\ve(\partial\L_\ve^k)$ at $\ve x$
and $R_\ve(t)$ has been defined in \eqref{eq:rest}.

We first estimate the third line in the formula above. Using Theorem
\ref{eq:EMYa} and Lemma \ref{eq:Dirichlet}, we get, for any $A>0$,
\begin{equation}
  \label{eq:11}
  \begin{split}
& \ve^{d-2} \int  \Big\{ \sum_{i=1}^d\sum_{x\in \L_\ve^k} 
\partial^{\ve,i}\l(t,\ve x)
\tau_x \V_i^\ve (\n) \Big\} \left( f_t - \wt{\Psi}_t^F  \right) (\n)
          d \nu\le\\
&\qquad\qquad\le A\ve^{d-4} D_0(\sqrt{f_t},\nu)+A\big(C_0'+C_0\big) +r(\ve,k,n,A)    
  \end{split}
  \end{equation}
where 
\[
\limsup_{n\to\infty}\limsup_{\ell\to\infty}\limsup_{\ve\to
  0}r(\ve,k,n,A)\le 0.
\]

We now examine the second line in \eqref{eq:6}. 
We can write $M_{x,i}^k$ as
$$
\d_iQ_{k,\ve}(x) + 2\ve\d_i\l(t,\ve x) \n^k(x) + const +
O(k^{-d})+O(\ve^3k^2)\, ,
$$
where 
\[
Q_{k,\ve}(\n,x)=\bigl(\n^k(x)-\e\lambda(t,\ve x)\bigr)^2
\]
and \textit{const} stands for a term independent of the
configuration.  By the entropy inequality, for any bounded function $G :\R
\times \pi^{d-1}\rightarrow \R$, for all $q>0$
\[
\begin{split}
\ve^{d-2}\int &\sum_{x\in\L_\ve^k} G(\ve x)
\wh{Q}_{k,\ve}(\n,x)
\left( f_t - \wt{\Psi}_t^F  \right) (\n) d \nu\\
& \le  q^{-1} \ve^{d-2} s_1(t) + q^{-1} \ve^{d-2} 
\log E^{\wt\Psi^F_t}\bigl[\exp q\sum_{x\in\L_\ve^k}\wh{Q}_{k,\ve}(\n,x)\bigr]\\  
\end{split}
\]
with
\[
\wh{Q}_{k,\ve}(\n,x):=Q_{k,\ve}(\n,x)- E^{\wt\Psi^F_t}[Q_{k,\ve}(\n,x)]\, .
\]
>From large deviations estimate
(Lemma 3.1.in \cite{EMY1}) there exists $q_0 >0$, such that, for
all $q< q_0$,
$$\lim_{\ell\to\infty}\limsup_{\ve\to 0}\ve^{d-2} 
\log E^{\wt\Psi^F_t}[\exp\, q\sum_{x\in\L_\ve^k} \wh{Q}_{k,\ve}(\n,x)]=0\, .$$
In conclusion, we got the following inequality for
the second line in \eqref{eq:6}
\[
\begin{split}
&\ve^{d-2}\int \sum_{i=1}^d\sum_{x\in\L_\ve^k} \partial^{\ve,i}\l(t,\ve x)
M_{x,i}^k(\n)
\left( f_t - \wt{\Psi}_t^F  \right) (\n) d \nu\qquad  \\
& \qquad \le \ve^{d-1}\int \sum_{i=1}^d\sum_{x\in\L_\ve^k}
\partial^{\ve,i}\l(t,\ve x) \big( 2\l(t,\ve x)\big)
\n^k(x)
\left( f_t - \wt{\Psi}_t^F  \right) (\n) d \nu  \\
& \qquad\ \ + q^{-1} \ve^{d-2} s_1(t) + r_{\ve,k}(q)\, ,
\end{split}
\]
where, for all $q< \frac{q_0}{2}$, $r_{\ve,k}(q)$ converges to 0
when $\ve \downarrow 0$ and $\ell \uparrow \infty$.

\smallskip To deal with the fourth line in \eqref{eq:6} (boundary term),
we need the following lemma
\begin{lem}
\label{eq:lem3.3}
Fix a bounded function $G: \R\times \pi^{d-1} \rightarrow \R$ and
$x_1\in [-\ve^{-1}+k,-\ve^{-1}+2k]\cup [\ve^{-1}-2k,\ve^{-1}-k]$. Then
there exists some constant $C>0$ such that for all 
$A>0$,
\[
\begin{split}
&\ve^{d-2}\int \sum_{y\in\pi_\ve^{d-1}} G(\ve (x_1,y))
\n^k(x_1,y)
\left( f_t - \wt{\Psi}_t^F  \right) (\n)
          d \nu \le \\
&\qquad\qquad\le \ve^{d-4}\frac{A}{2}\left(D_b\big(\sqrt{h_t},\wt{\Psi}_t \nu \big)     +D_0(\sqrt{f_t},\nu)\right)
+ \ve k C  \, \| G\|_\infty  \left( \frac{\| G\|_\infty}{A} +1 \right)\, . 
\end{split}
\]
\end{lem}
The proof of this lemma is postponed to the next section.
\bigskip

We now turn to the proof of Proposition \ref{eq:prop2} and we first
consider the term $E_2$. Since the function $(\partial_t \l)$ is equal to 0
outside $\Omega_\t$ (cf. formula \eqref{eq:7} and below),
a simple computation shows that
\[
\begin{split}
E_2  &= -\ve^{d-2} \int\frac{\partial_t \wt{\Psi}_t^F}{\wt{\Psi}_t^F} f_t\, d\nu\\
 & =-\ve^{d-1}\int \sum_{x\in \L_\ve}
   \partial_t (\l *\omega)(t,\ve x)\n(x)
  \left( f_t - \wt{\Psi}_t^F  \right) (\n) \, d \nu \\
&=-\ve^{d-1}\int \sum_{x:\ve x\in \Omega_\t} \partial_t\l (t,\ve x)\n^k(x)
  \left( f_t - \wt{\Psi}_t^F  \right) (\n) \, d \nu \, .
\end{split}
\]

To conclude the proof of proposition \ref{eq:prop2},
we integrate the inequality
(\ref{eq:3.9}) from 0 to $t$. Combining with the above estimates, 
we obtain that there exist constants
$K_1>0$, $K_2>0$, $K_3>0$ and $c>0$ such that, for all small enough $A>0$,
\begin{equation}
\label{eq:3.10}
\begin{split}
\ve^{-2} s_1(t) & \le K_1\ve^{-2}\int_0^t s_1(u) du \\
\ & \ \ -K_2 \ve^{d-4}\int_0^t ds
\left\{ D_0\big(\sqrt{h_s^F},\wt{\Psi}_s^F \nu \big) 
+D_b\big(\sqrt{h_s},\wt{\Psi}_s \nu \big)  \right\}\\
\ &\ \  +\ve^{d-1}\int_0^t ds \Big\{
 \sum_{x\in\Omega_{\ve k}}H(s,\ve x)
\int \eta^k(x) (f_s -\wt{\Psi}_s^F) d\nu \Big\}\\
 \ & \ \  +A K_3\ve^{d-4}\int_0^t 
 D_0\big(\sqrt{f_s}, \nu \big)  ds+ r(\ve ,k,n,A)+cA\, ,
\end{split}
\end{equation}
where $\displaystyle\limsup_{n\to\infty}\limsup_{\ell\to\infty}\limsup_{\ve\to
  0}r(\ve,k,n,A)\le 0$ for all $A>0$, and the function $H$ is given by
$$
H(s, \ve x) =\sum_{i=1}^d \delta_i \partial_{e_i}\big( \l(s,\ve x)\big)^2
+\sum_{1\le i,j\le d}D_{i,j}\partial^2_{u_i,u_j}\l(s,\ve x) 
-\partial_s \l (s,\ve x)\, .
$$
Recalling that $\l(t,u)$ satisfies
\eqref{eq:8}  in $\Omega_{2\t}$ and using the notation $R_\ve$
defined in \eqref{eq:rest}, we have
$$
\ve^{d-1}
 \sum_{x\in\Omega_{\ve k}}H(s,\ve x)
\int \eta^k(x) (f_s -\wt{\Psi}_s^F) d\nu
=\ve^{d-1}
 \sum_{x\in\Omega_{\ve k}\setminus\Omega_{2\t}}H(s,\ve x)
\int \eta^k(x) (f_s -\wt{\Psi}_s^F) d\nu+R_\ve(s).
$$
We will prove in the next section:
\begin{lem}
\label{eq:l-bord}
For any bounded function $G$ defined on $\L$ and for any density $f$ with
respect to $\nu$, we have for any $A>0$
\[
\begin{split}
&\ve^{d-1} \sum_{\ve x\in \Omega_{\ve k}\setminus \Omega_{{2\theta}} } G(\ve x) \int
 \eta^k(x) \left( f_s - \wt{\Psi}_s^F  \right) (\n)\,d \nu\\
&\ \ \le A \ve^{d-4} \left\{ D_b\bigl(\sqrt{h_s},\wt{\psi}_s\nu \bigr) 
+D_0\bigl(\sqrt{f_s},\nu\bigr)\right\} +C  \t^3\|G \|_\infty
\left(\frac{\|G \|_\infty}{A}+1\right)
\end{split}
\]
for some positive constant $C$.
\end{lem}
Therefore, noticing that $\|H\|_\infty\le\mathrm{const}\,\t^{-2}$, we
get for any small enough $A>0$
\begin{equation}
\label{eq:3.10bis}
\begin{split}
\ve^{-2} s_1(t) & \le K_1\ve^{-2}\int_0^t s_1(u) du \\
\ & \ \ -K_2 \ve^{d-4}\int_0^t \!ds\,
\left\{ D_0\big(\sqrt{h_s^F},\wt{\Psi}_s^F \nu \big) 
+D_b\big(\sqrt{h_s},\wt{\Psi}_s \nu \big)  \right\}\\
 \ & \ \  +A K_3\ve^{d-4}\int_0^t\!ds\,
 D_0\big(\sqrt{f_s}, \nu \big)
+ r(\ve ,k,n,A)+c\left(A+\frac{\t}{A}\right)\, ,
\end{split}
\end{equation}
where the constants previously defined may have changed their values.
\medskip



To conclude, we use Lemma \ref{eq:Dirichlet} to control the
Dirichlet form $\ve^{d-4}D_0(\sqrt{f_s},\nu)$ with
$\ve^{d-4}D_0(\sqrt{h_s^F},\wt{\Psi}_s^F \nu )$. The error obtained
from this replacement is a constant but we notice that there is the
small factor $A$ in front of this term. Therefore there exists a constant
$c'>0$ such that for any small enough $A>0$,
\begin{equation*}
  \begin{split}
    \ve^{-2} s_1(t) & \le K_1\ve^{-2}\int_0^t s_1(u) du \\
\ & \ \ -K_2 \ve^{d-4}\int_0^t \!ds\,
\left\{ D_0\big(\sqrt{h_s^F},\wt{\Psi}_s^F \nu \big) 
+D_b\big(\sqrt{h_s},\wt{\Psi}_s \nu \big)  \right\}\\
 \ & \ \ + r(\ve ,k,n,A)+c'\left(A+\frac{\t}{A}\right)\, ,
  \end{split}
\end{equation*}
then we choose $A=A(\t)$ vanishing with $\t$ in such a way that
$\displaystyle\lim_{\t\to 0}\t A^{-1}(\t)=0$ (e.g. $A(\t)=\sqrt{\t}$).
Finally we consider the successive limits $\ve\to 0$, $\ell\to
\infty$, $n\to \infty$, $\t\to 0$ and we apply Gronwall lemma.

\cqfd
\section{Estimates on boundary terms}

\bigskip
\noindent{\bf Proof of Lemma \ref{eq:rest-local}.} 
 From the definition of  $\wh{{W}}_{x,i}$ we have that the term $E_1^4$
 in
(\ref{eq:localisation}) can be written as
$$\ve^{d-2} \int \left\{ \sum_{i=1}^d \sum_{x\in\Gamma_\ve^k} 
H^{\ve,i} (t,\ve x)
 \frac{1}{|{\L}_k|}\sum_{y\in{\bar\L}_k(x)}
  \wh{{W}}_{y,i} \right\}
\big( f_t - \wt{\Psi}_t^F \big)(\n) d \nu$$
where ${\bar\L}_k(x)$ is the block of rectangular shape which is the
set of $y\in {\L}_k(x)$ (cube centered in 
$x$) such that $y$ and $y+e_1$ belong to $\L_\e$.  
Let $\bar M_{x,i}^k(\n)$ be the conditional expectation of 
${{W}}_{x,i}, x\in \Lambda_\e$ 
given the density of particles on  $\bar{\L}_{k}(x)$.
Repeating the argument given in the paragraph following inequality
\eqref{eq:11}, it suffices to prove that,
for all $A>0$, for any bounded function $J$ and any probability
density 
$f$ with respect to $\nu$,
\[
\begin{split}
\lim_{n\to \infty}\lim_{\ve \rightarrow 0}\Bigg\{
\ve^{d-2} \int \Big( \sum_{i=1}^d \sum_{x\in\Gamma_\ve^k} J(\ve x)\Big[
 \big(\wh{\overline{W}}_{0,i}*\omega\big)(x)
&-\bar M_{x,i}^k(\n)\Big]
\Big)
 f d \nu 
 \\
&\qquad -A \ve^{d-4} D_0(\sqrt{f})  \Bigg\}
   =0\, .
   \end{split}
\]

We need the following definition:

Let $\mu_\ell$ be the canonical measure in the block $\Lambda_\ell$
with given density $\eta^\ell$. For any $\{x,y\}\subset \L_\ell$, we
introduce the Dirichlet form $D^b_0(h)$ as
\[
D^{\{x,y\}}_0(h)=\int\!r_{\{x,y\}}(\n)\bigl(h(\n^{\{x,y\}})-h(\n)\bigr)^2\,d\mu_\ell
\]
and we define the finite volume variance
$$
V _\ell (G,\eta^\ell) = ( 2 \ell + 1 ) ^{-d} \langle \sum _{|x|
  \leq \ell} ( \tau _x G - E[G|\eta^\ell]) ( - \cl^{(s)}_{\ell} )
^{-1} \sum _{|x|\leq \ell} ( \tau _x G - E[G|\eta^\ell]) \rangle _{\mu
  _{\ell}}.$$
where $\cl^{(s)}_{\ell}$ is the symmetric part of the
generator $\cl_{\e,0}$ restricted to the box $\L_\ell$.
\noindent The proof of Theorem 4.6 in
\cite{EMY1} is based on the following result:

\begin{lem}
\label{emy}
For any cylinder function $h$ there exist a constant $C$ and a function $C(h,\ell)$ vanishing for
$\ell\to\infty$ such that for any positive $A$ and $\bar\ell=\ell^{d+2}$
\begin{equation*}
  \begin{split}
\int \omega*[\tau_x h- E[h|\eta^k(x)]f_td\mu_k - &A
\frac{1}{|\Lambda_k|}\sum_{\{y,z\}\subset\Lambda_k(x)}
D_0^{\{y,z\}}(\sqrt{f})\\
&\qquad\le\frac{C}{A} \e^2\int
V_{\bar\ell}(h,\eta^{\bar\ell})f_td\mu_k + \e^2 C(h,\ell)   
  \end{split}
\end{equation*}
\end{lem}

The proof of this Lemma is given in \cite{EMY1} for square blocks but
it can be extended easily to a rectangular shape provided that the
volume of the block is of order $k^d$.
Using  Lemma \ref{emy} we prove Lemma \ref{eq:rest-local} by taking 
the expectation with respect to $\nu$, multiplying by
$\e^{d-2} J$,  summing over 
$x\in
\Gamma^k_\e$ and noting that the number of terms in $\Gamma^k_\e$ is $\e^{-d+1}k$.

\bigskip
\noindent{\bf Proof of Lemma \ref{eq:lem3.3}.}
Fix $\ve^{-1}-2k \le x_1\le \ve^{-1}-k$ and write the sum as
\[
\begin{split}
&\ve^{d-2}\int \sum_{y\in\pi_\ve^{d-1}} G(\ve(x_1,y))\n^k(x_1,y)
(f_t -\wt{\Psi}_t^F)d\nu\\
&\quad =\frac{\ve^{d-2}}{k}\sum_{|z_1|\le k}
\int \sum_{y\in\pi_\ve^{d-1}} \big(G(\ve(x_1,y))\big)^k\n(x_1+z_1,y)
(f_t -\wt{\Psi}_t^F)d\nu\, ,
\end{split}
\]
where $\big(G(\ve(x_1,y))\big)^k$ is the $(d-1)$-dimensional convolution in the variable $y\in\pi_\ve^{d-1}$. Therefore, to prove the lemma, it is 
enough to show that, if $H :\R\to \R$ is a bounded function, then for every $\ve^{-1}-3k \le z_1\le \ve^{-1}$, and $A>0$
\[
\begin{split}
&\ve^{d-2}
\int \sum_{y\in\pi_\ve^{d-1}} H(\ve(z_1,y))\n(z_1,y)
(f_t -\wt{\Psi}_t^F)d\nu\\
&\ \ \ \ 
 \le \ve^{d-4}\frac{A}{2}\left(D_b\big(
\sqrt{h_t},\wt{\Psi}_t \nu \big)     +D_0(\sqrt{f_t},\nu)\right)
+ \textrm{const}\,\  \ve k\|G \|_\infty
\left(\frac{\|G \|_\infty}{A}+1\right)
\, , 
\end{split}
\]

Let $\ve^{-1}-3k \le z_1\le \ve^{-1}$
and decompose the left hand side of the last inequality into
two terms $B_1$ and $B_2$ 
\[
\begin{split}
&\ve^{d-2}
\int \sum_{y\in\pi_\ve^{d-1}} H(\ve(z_1,y))
   \left( \n(z_1,y) -\n(\ve^{-1},y)\right)
(f_t -\wt{\Psi}_t^F)d\nu\\
&\ \ +\ve^{d-2}
\int \sum_{y\in\pi_\ve^{d-1}} H(\ve(z_1,y))
   \n(\ve^{-1},y)
(f_t -\wt{\Psi}_t)d\nu\\
&:=B_1 +B_2\, . 
\end{split}
\]

The term $B_2$ is the simplest one. From an integration by parts 
and Schwartz inequality, it is bounded, for all $A>0$, by
\begin{equation}
\label{eq:l1.1}
B_2 \le
  \ve^{d-2}   \sum_{y\in\pi_\ve^{d-1}} \left\{ \frac{A}{2} \ve^{-2} 
  D_{(\ve^{-1},y)}^b\big(\sqrt{h_t},\wt{\Psi}_t \nu) +\ve^2
\frac{\mathrm{const}}{A} 
\| H\|_\infty^2  \right\} \, ,
\end{equation}
where for a function $f$, a positive measure $\mu$ and $x\in \Gamma_\ve$
$$
D_x^b(f,\mu)=\left< \left( f(\n^{x}) -f(\n)\right)^2 \right>_\mu.
$$

We now consider $B_1$. By Schwartz inequality 
\[
\begin{split}
B_1
& =\ve^{d-2}
\int \sum_{y\in\pi_\ve^{d-1}}\sum_{z=z_1}^{\ve^{-1}-1} H(\ve(z_1,y))
   \left( \n(t+z+1,y) -\n( t+z ,y)\right)
(f_t -\wt{\Psi}_t^F)d\nu\\
& \le \ve^{d-2}  \left\{ \frac{\ve^{-2} A}{2}  
  D_0\big(\sqrt{f_t}, \nu) +\frac{\ve^{-d+3} k}{2A}
\| H\|_\infty^2  \right\}
+ 
\ve k C\,\|
  H\|_\infty   
\end{split}
\]
for some constant $C>0$,
where we have used in the last inequality, for the second term corresponding to the integration with
respect to $\wt{\Psi}_t^F\nu$, integration by parts and Taylor
expansion.
\cqfd

\bigskip
\noindent{\bf Proof of Lemma \ref{eq:l-bord}.}
The summation over the set $\left\{x\ ,\ \ve x\in \Omega_{\ve k}\setminus \Omega_{2\t}\right\}$ can be
divided in two similar terms. We consider the one where the first
coordinate is such that $\ve^{-1}(1-2 \t)\le  x_1 \le \ve^{-1}-k$,
the second term is handled in the same way.
If we repeat the arguments used in the proof of the Lemma
\ref{eq:lem3.3}, we obtain for $\ve^{-1}(1-2 \t)\le  x_1 \le \ve^{-1}-k$,
\[
\begin{split}
&\ve^{d-1}\sum_{y\in\pi_\ve^{d-1}}
G(\ve (x_1,y)) \int \n^k (x_1,y) \left( f_t(\n)-\wt{\Psi}_t\right)
d\nu \\
&\ \ \ \le \ve  \left\{
   \ve^{d-4} \frac{A}{2\t} \left( D_0\big(\sqrt{f_t}, \nu \big) +
    D_b \big(\sqrt{h_t}, \wt{\Psi}_t\nu  \big)\right)
  + C \t^2\| G\|_\infty \Bigl(\frac{\| G\|_\infty}{A} +1\Bigr))
                \right\}\, .
\end{split}
\]
To  conclude the proof of the lemma, we just have to take the sum over $x_1$.
\cqfd

\bigskip
\section{Comments}
We conclude with a few generalizations:

\noindent 1) Assumption $\delta_1>0$. 

\noindent If $\delta_1=0$ we need to introduce  a  different boundary generator to fix the
density on the boundary. We make the choice which is usually done for the symmetric  case, a death and birth process $\bar
{ \cl }_b$ acting on each site of the  boundary such that  it is reversible with respect to the one site measure with density
$\rho(\ve x)=1/2+\ve b(\ve x)$
\[
\bar\cl_bf(\n)=\sum_{x\in\Gamma_\ve}\Big[\rho(\ve x)(1-\n_x)
\left[f(\n^{x,+})-f(\n)\right]
+(1-\rho(\ve x))\n_x
\left[f(\n^{x,-})-f(\n)\right]\Big]
\]
where $\n^{x,+}$ and $\n^{x,-}$ are the configurations obtained from
$\n$ by adding or removing a particle at site $x$. 
\bigskip

\noindent 2) General domain.

\noindent We generalize  now the model to the case of a macroscopic system in a smooth bounded convex domain of
$\R^d$. We introduce the  boundary generator acting on the boundary $\Gamma_\ve$  as in  Section 2  with the boundary rates
$C^b(\ve x,\eta)$ chosen in such a way that a particle is added at the site
$x$ on the boundary when the site is empty, with rate $|\d\cdot n| (1/2 +\ve b(\ve
x))$ if  $\d\cdot n$, the scalar product of the vector $\delta$ and the outward normal in $x$, is positive and 
  removed when the site is occupied, with rate $|\d\cdot n|(1/2 -\ve b(\ve x))$ if the scalar product is negative. For $x\in
\Gamma_\ve$ 
\[
  \begin{split}
C^b (\ve x, \n)      &= |\d\cdot n|
\bigl(\frac{1}{2}+\ve\  b (\ve x)\bigr)(1-\n (x))\1\{\d\cdot n(\ve x)>0\} \\&+|\d\cdot n|
\bigl(\frac{1}{2}-\ve\  b (\ve x) \bigr) \n (x) \1\{\d\cdot n(\ve x)<0\}.
\end{split}
\]
These rates  fix the
value of the particle density on the macroscopic  boundary to be $1/2+\ve\  b(x)$. This choice is sufficient to remove  the
entropy flow generated by the drift of ASEP. Since $\d\cdot n(\ve x)$ can be zero in some points for a general domain we have
to add, on the
basis of the previous remark, also a reversible boundary generator to fix the density in these sites. 

\bigskip

\noindent 3) Different boundary conditions.

\noindent To remove the entropy flux it is possible to choose a  generator $\bar{\cl _b}$ which is reversible instead of using
the non reversible  generator
$\cl_{\e,b}$, but then we have to speed up it by a factor $\e^{-3}$ with respect to  the jump process. The
total generator speeded up is then
\[
\ve^{-2}\cl_\ve=\ve^{-2}\cl_{\ve,0}+\ve^{-5}\bar\cl_{b}.
\]
We note that the bad boundary terms in our proof are eliminated by means of a cancellation. If we use only the reversible
generator $\e^{-2}\bar{\cl _b}$ on the boundary they will be controlled by the Dirichlet form associated to
$\bar\cl_{b}$, by a generalization of the Lemma 3.3.
\vskip1cm
{\it Acknowledgments}. O.B. and  M.M. would like to thank the 
hospitality of the University of Roma Tor Vergata and  R.E and R. M. the hospitality of the
University of Rouen. This work has been partially supported by GNFM-INdAM and MURST.

\vskip1cm

\end{document}